\documentclass[twocolumn,letter]{jpsj3}
\usepackage{txfonts}
\usepackage{color}

\title{Magnetization and Magnetoresistance of CeRu$_2$Al$_{10}$\\ 
 under High Magnetic Fields along $c$-Axis }

\author{Akihiro \textsc{Kondo}$^{1}$\thanks{E-mail address: kondo@issp.u-tokyo.ac.jp}, Junfeng \textsc{Wang}$^{1, 2}$, Koichi \textsc{Kindo}$^{1}$,\\
Tomoaki \textsc{Takesaka}$^{3}$, Yuta \textsc{Ogane}$^{3}$, Yukihiro \textsc{Kawamura}$^{3}$, Takashi \textsc{Nishioka}$^{3}$, \\
Daiki \textsc{Tanaka}$^{4}$, Hiroshi \textsc{Tanida}$^{4}$, and Masafumi \textsc{Sera}$^{4}$}

\inst{$^{1}$Institute for Solid State Physics, University of Tokyo, Kashiwa, Chiba 277-8581, Japan\\
$^{2}$Wuhan High Magnetic Field Center, Huazhong University of Science and Technology, Wuhan 430074, China \\
$^{3}$Graduate school of Integrated Arts and Science, Kochi University, Kochi 780-8520, Japan\\
$^{4}$Department of Quantum Matter, AdSM, Hiroshima University, Higashi-hiroshima 739-8530, Japan}

\abst{We have studied the magnetization and magnetoresistance of CeRu$_2$Al$_{10}$ in the applied magnetic field $H$ along the $c$-axis up to $\sim55$ T. 
The magnetization $M$ at low temperatures shows an $H$-linear increase with a small slope of $M$/$H$ than that for $H\parallel$ $a$-axis up to $\sim55$ T after showing a small anomaly at $H^*\sim4$ T, which indicates that the critical field to the paramagnetic phase  $H_c^p$ is higher than 55 T for $H\parallel$ $c$-axis. 
The magnetization curves for $H\parallel a$- and $c$-axes below the antiferro magnetic (AFM) transition temperature $T_0$ behave as if the magnetic anisotropy in the AFM-ordered phase is small, although there exists a large magnetic anisotropy in the paramagnetic phase, which favors the easy magnetization axis along the $a$-axis. 
On the other hand, very recently, Khalyavin $et\ al$. have reported that the AFM order where the magnetic moment is parallel to the $c$-axis takes place below $T_0$. 
These results indicate that the AFM order in this compound is not a simple one. 
The longitudinal magnetoresistance for $H\parallel c$-axis at low temperatures shows no anomaly originating from the phase transition, but shows oscillations below 4.2 K. 
This oscillatory behavior below 4.2 K originates from the Shubnikov-de Haas oscillations, from which the cross section of the Fermi surface normal to the $c$-axis is estimated to be $\sim1.0\times 10^{14}$ cm$^{-2}$, with no large effective mass.
This is the first direct evidence of the existence of the Fermi surface below $T_0$.   
 }

\kword{CeRu$_2$Al$_{10}$, Kondo semiconductor, pulsed high magnetic field}

\begin{document}
\maketitle

Recently, a new ternary compound CeRu$_2$Al$_{10}$, with an orthorhombic YbFe$_2$Al$_{10}$-type structure\cite{st1, st2}, has attracted much attention because of its novel phase transition at $T_0$ = 27 K. \cite{Stry, Nishi, Matsu}
The similar transition is observed also in CeOs$_2$Al$_{10}$.\cite{Nishi}
In the first report by Strydom, the long-range order (LRO) below $T_0$ was proposed to be an antiferro magnetic (AFM) order. \cite{Stry}
Nishioka $et\ al$. studied Ce$T_2$Al$_{10}$ ($T$=Ru, Os, Fe) single crystals and proposed the charge density wave (CDW) transition\cite{Nishi}. 
Matsumura $et\ al$. performed an $^{27}$Al NQR experiment and proposed that  the phase transition originates not from the magnetic one but  from the nonmagnetic one.\cite{Matsu} 
Tanida $et\ al$. examined La substitution and the magnetic field effect on Ce$_{x}$La$_{1-x}$Ru$_2$Al$_{10}$ and found that the origin of the LRO is the magnetic interaction between Ce ions; they proposed the singlet pair formation between Ce ions as the origin of the LRO. \cite{Tany1}  
They also pointed out that the system may have a two-dimensional character in the $ac$-plane. \cite{Tany2}
Hanzawa theoretically investigated the singlet ground state formed by the dimer of the two Ce ions with a large magnetic anisotropy in the crystalline electric field (CEF) doublet ground state\cite{Han}.

In our previous study, we performed the magnetization measurements of Ce$_{x}$La$_{1-x}$Ru$_2$Al$_{10}$ ($x$ = 1, 0.75) under high magnetic fields for the applied magnetic field $H$ along the $a$-axis. \cite{Kon}
We successfully obtained the magnetic phase diagram of these two compounds for $H\parallel$ $a$-axis, and found that the LRO for $x$ = 1 disappears at $H_c^p\sim50$ T, which is the critical field for the paramagnetic phase. 
We also pointed out the possibility of the appearance of a magnetic-field-induced phase between $\sim40$ and $\sim50$ T, although its origin  is not known at present.
For $x$ = 0.75, $H_c^p$ decreases to $\sim37$ T. 
The magnetic phase diagram and magnetization curve are qualitatively consistent with Hanzawa's mean field calculation results. \cite{Han} 

Quite recently, Khalyavin $et\ al$. have reported that the LRO is a collinear AFM order with a propagation vector $\bf{k}$ = (1,0,0).\cite{Kha} 
Here, we summarize the experimental results.  
\begin{description}
\item[(1)] $T_0\sim27$ K is very high among the series of $R$Ru$_2$Al$_{10}$ ($R$ = rare-earth element) as the AFM order considering that,  even in GdRu$_2$Al$_{10}$, $T_{\rm N}=16.5$ K\cite{Nishi}. 
\item[(2)] The magnetic susceptibility $\chi$ shows a large anisotropy $\chi_a$$>$$\chi_c$$>$$\chi_b$ above $T_0$. $\chi$ shows a decrease below $T_0$ along all the crystal axes.\cite{Nishi} 
In addition, a large Van Vleck term exists below $T_0$ at least for $H\parallel c$-axis. 
 In this case, a finite magnitude of $\chi_c$ as large as 65\% remains at low temperatures when compared with $\chi_c$ at $T_0$, although the AFM moment ($M_{\rm AF}$) is along the $c$-axis.\cite{Kha} 
 Here, we conveniently use the word "Van Vleck" in order to represent the following characteristics; the magnetization ($M$) curve shows an $H$-linear increase and $M$ is temperature-independent at low temperatures.  
\item[(3)] The magnetization curve for $H\parallel a$-axis shows an almost $H$-linear increase up to $\sim30$ T and a concave increase up to $H_c^p\sim50$ T at 1.3 K. 
Above $H_c^p$, $M$ shows a small increase, which mainly comes from the usual Van Vleck contribution from the excited CEF state. 
\item[(4)] The magnetization curve shows an anomaly at $H^*\sim4$ T only for $H\parallel c$-axis.\cite{Tany3} 
\item[(5)] A small internal field is observed below $T_0$ by the $\mu$SR experiment. \cite{mSR}
\item[(6)] The elastic neutron scattering experiment indicates an AFM order with propagation vector $\bf{k}$ = (1,0,0).\cite{Kha} 
$M_{\rm AF}$ with $\sim0.34\mu_{\rm B}$/Ce is parallel to the $c$-axis. 
In this case, when the magnetic field is applied along the $c$-axis, a spin-flop or metamagnetic transition is expected to appear. 
\item[(7)] The inelastic neutron scattering experiment reveals the existence of a large spin gap with an excited energy centered at $\sim8$ meV, which develops below $T_0$. \cite{ND}
\item[(8)] The nuclear-spin lattice relaxation rate 1/$T_1$ of the $^{27}$Al NQR measured at the Al(5) site indicates the existence of a gap of $\sim100$ K below $T_0$.\cite{Matsu} 
\item[(9)] The specific heat indicates the existence of a gap with $\sim100$ K below $T_0$, accompanied by a large magnetic entropy. 
\cite{Nishi, Tany1}
\item[(10)] The volume of CeOs$_2$Al$_{10}$ deviates towards a smaller value from that expected for the Ce$^{3+}$ ion in the series of $R$Os$_2$Al$_{10}$, indicating that CeOs$_2$Al$_{10}$ contains a valence fluctuating character.\cite{st1} On the other hand, the valence of the Ce ion in CeRu$_2$Al$_{10}$ is close to +3. \cite{Tany1}
\item[(11)] The electrical resistivity shows a Kondo semiconducting behavior at high temperatures,\cite{Stry, Nishi} which is also observed in CeOs$_2$Al$_{10}$.\cite{Nishi, Muro2}
\end{description}
Although it is confirmed that the AFM order takes place below $T_0$ from results (5) and (6),  its the consistency with the macroscopic properties has never been discussed,and thus the origin of the LRO in CeRu$_2$Al$_{10}$ remains controversial. 

In order to understand the unusual LRO in this system, it is necessary to clarify the overall features of the anisotropic phase diagram. 
In our previous study\cite{Kon}, we performed the magnetization measurement for $H\parallel$ $a$-axis, because the lowest critical magnetic field among the three crystal axes was expected to be along the $a$-axis. 
The purpose of the present study is to determine how high is the critical magnetic field for  $H\parallel$ $c$-axis. 
If the critical field $H_c^p$ could be observed for $H\parallel$ $c$-axis, we can obtain much information on the CEF ground state from the magnitude of the saturation magnetization for $H\parallel$ $c$- and $a$-axes. 
Furthermore, we performed the magnetoresistance measurement at high magnetic fields in order to obtain information on the scattering mechanism of conduction electrons and the electronic properties of the ground state. 
Note that we report here the first observation of the Shubnikov-de Haas (SdH) oscillations in this system.

The single crystals of CeRu$_2$Al$_{10}$ used in the present study were prepared by the Al self-flux method.
Pulsed magnetic fields up to 55 T were generated with a duration of 36 ms using nondestructive magnets. 
Magnetization was measured by the induction method using a standard pick-up coil in magnetic fields along the $c$-axis.  
The absolute value was calibrated by comparing the data with the magnetization data below 7 T, measured using MPMS (Quantum Design). 
Magnetoresistance was measured by a standard four-probe dc method. 
The electrical current flows along the $c$-axis. 

\begin{figure}[tb]
\begin{center}
\includegraphics[width=.85\linewidth]{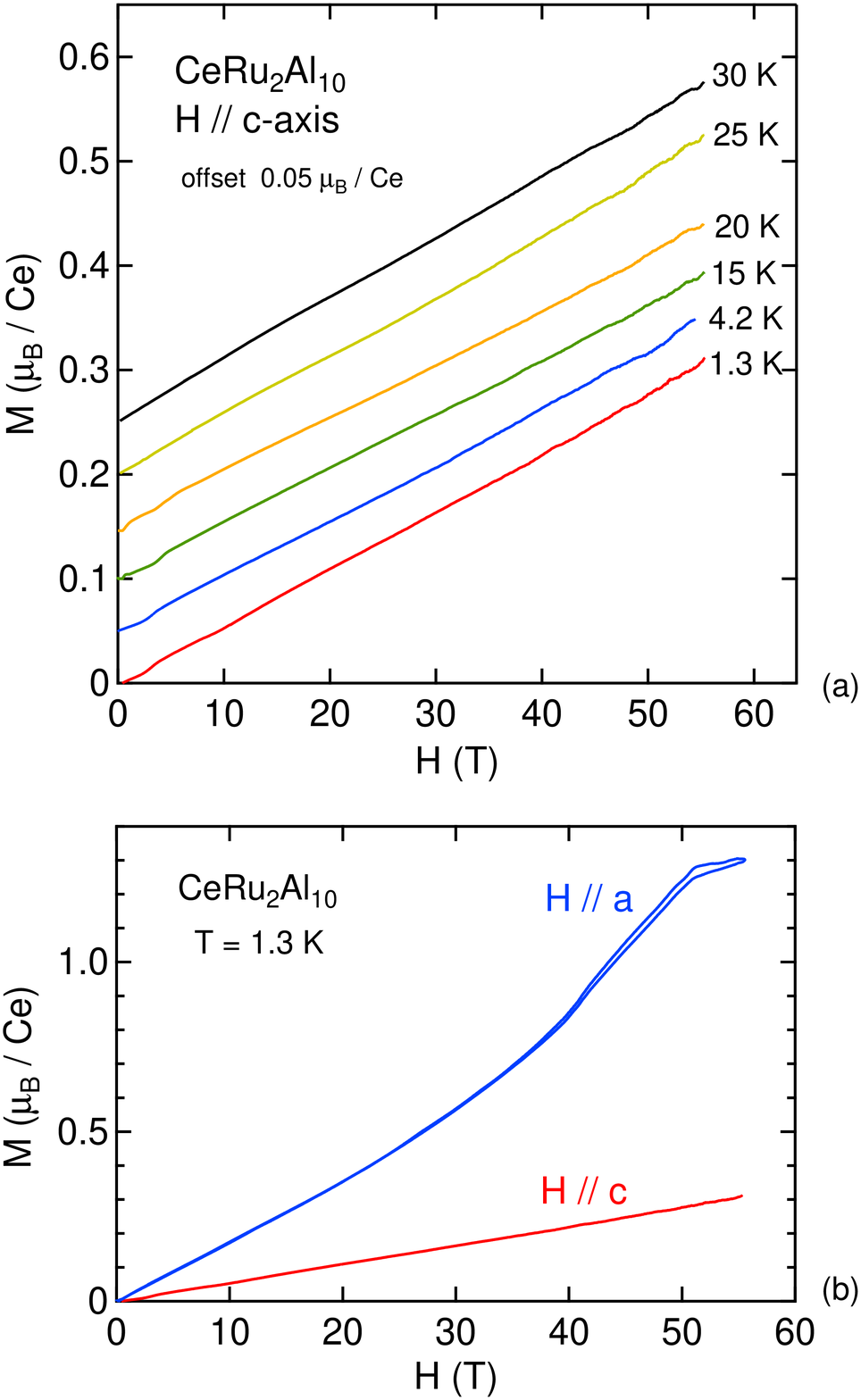}
\caption{(Color online)  (a) Magnetic field dependence of the magnetization of CeRu$_2$Al$_{10}$ at various temperatures for $H\parallel c$-axis. (b) Magnetization curve of CeRu$_2$Al$_{10}$ at $T$ = 1.3 K for $H\parallel a$- and $c$-axes. The data for  $H\parallel a$-axis is cited from ref. 11.}
\label{Fig1}
\end{center}
\end{figure}

\begin{figure}[tb]
\begin{center}
\includegraphics[width=.85\linewidth]{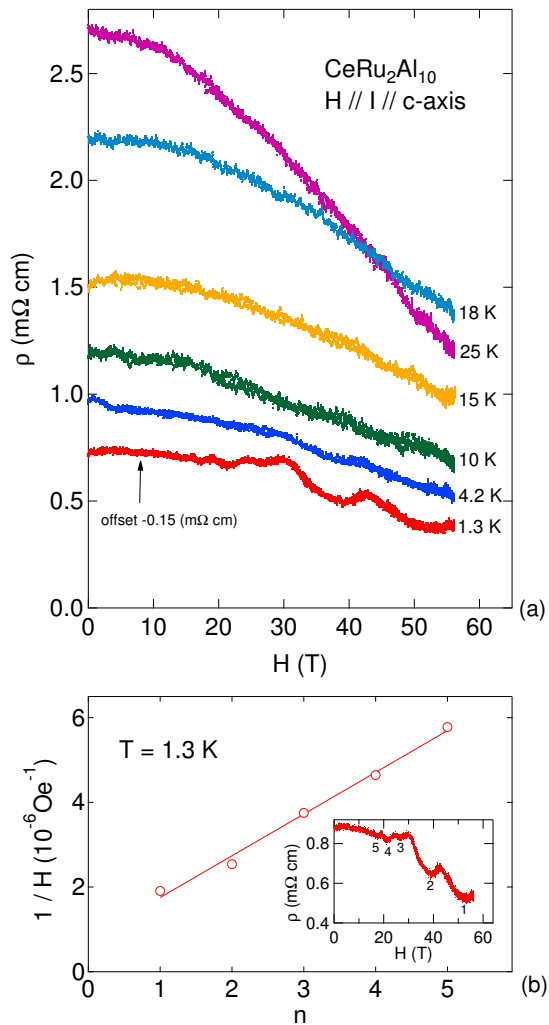}
\caption{(Color online) (a) Longitudinal magnetoresistance of CeRu$_2$Al$_{10}$ under various temperatures for $H\parallel c$-axis. The origin of the vertical axis for $T$ = 1.3 K is shifted for easy viewing. (b) Inverse of $H$, which corresponds to a local minimal value of the magnetoresistance versus $n$. $n$ denotes an integer corresponding to each minimum of oscillations as shown in the inset. }
\label{Fig2}
\end{center}
\end{figure}

Figure 1(a) shows the $H$ dependence of the magnetization ($M$) of CeRu$_2$Al$_{10}$ for $H\parallel$ $c$-axis. 
In Fig. 1(b), we show the $M$-$H$ curves at 1.3 K for $H\parallel a$- and $c$-axes. 
For $T\leq20$ K, a small anomaly is observed at 4 T, which corresponds to $H^*$ reported by Tanida $et\  al$. \cite{Tany3}
Above $H^*$, $M$ shows an $H$-linear increase with a much smaller slope than for $H\parallel a$-axis. 
For $H\parallel$ $a$-axis, two anomalies accompanied by a phase transition are observed at $\sim40$ and $\sim50$ T.\cite{Kon} 
On the other hand, for $H\parallel$ $c$-axis, there is no anomaly such as metamagnetic transition at high magnetic fields expected for $H\parallel$ $c$-axis if this compound is a simple AF magnet with a large magnetic anisotropy. 
The  magnitude of $M$ at $\sim55$ T for 1.3 K is $\sim$0.3$\mu_{\rm B}$/Ce, which is three times smaller than $\sim1.3\mu_{\rm B}$/Ce at $H_c^p$ for $H\parallel$ $a$-axis. 
These results suggest that $H_c^p$ for $H\parallel$ $c$-axis is larger than that for $H\parallel$ $a$-axis. 
Although the magnetization measurement for $H\parallel$ $b$-axis has not yet been performed, $H_c^p$ for $H\parallel$ $b$-axis may be much larger than that for $H\parallel$ $c$-axis because the $b$-axis is the magnetization hard axis. 

Figure 2(a) shows the longitudinal magnetoresistance $\rho$ of CeRu$_2$Al$_{10}$ for $H\parallel$ $c$-axis. 
Although the magnetic field dependence of $\rho$ is small at low magnetic fields below $\sim20$ T at low temperatures, $\rho$ exhibits a negative magnetoresistance at high temperatures. 
The latter is due to the suppression of the conduction electron scattering by the localized magnetic moments. 
$\rho$ shows no anomaly accompanied by a phase transition, but shows oscillations at 4.2 K. 
With decreasing temperature, five oscillations are clearly recognized at 1.3 K. 
In order to check if this oscillatory behavior is intrinsic or not, we performed the magnetoresistance measurement in a different maximum pulsed field up to $\sim45$ T and confirmed the reproducibility. 
The oscillatory behavior below 4.2 K is expected to originate from the SdH oscillations. 
In Fig. 2(b), we plot the inverse of $H$, 1/$H$, as a function of $n$ at 1.3 K. 
Here, $n$ denotes an integer corresponding to each minimum of oscillations as shown in the inset of Fig. 2(b). 
1/$H$ is approximately linear to $n$. 
This result means that $\rho$ oscillates periodically in 1/ $H$. 
Thus, we concluded that the origin of the oscillatory behavior of $\rho$ at low temperatures originates from the SdH oscillations. 
From the observed oscillations, the cross section of the Fermi surface normal to the $c$-axis is estimated to be $\sim1.0\times 10^{14}$ cm$^{-2}$. 
The effective mass, $m^*$ is very roughly estimated to be $\sim2\ m_0$ from the results at 1.3 and 4.2 K. 
Here, $m_0$ is the free electron mass. 
Since the sign of the Hall resistivity is positive\cite{Tany2}, hole carriers may dominate the electrical conduction. 
These suggest that the observed Fermi surface is that of a hole band. 
Here, we note that the result in the same configuration up to 14.5 T shows a positive magnetoresistance, which is different from the present negative one. 
At present, we do not know the exact reason for this difference, although such possibilities that the samples were not the same and that there exists a small misalignment of the sample setting between two experiments could be considered. 
However, as far as the oscillation of $\rho$ in magnetic field, by considering the 1/$H$ periodic dependence and a rapid smearing out of the oscillation with increasing temperature, it is considered that the oscillation of $\rho$ is the SdH one. 

Here, we discuss the high-field magnetization results of CeRu$_2$Al$_{10}$. 
{\it We would like to insist that the AFM order in CeRu$_2$Al$_{10}$ is not a simple AFM one.  }
It is confirmed that the AFM order takes place below $T_0$ from the above-mentioned results (5) and (6). 
However, as we discuss below, it is difficult to explain the experimental results by the simple AFM order proposed by Khalyavin $et\ al$.\cite{Kha}
First, we note the problem of this simple AFM order. 
The AF magnetic moment $M_{\rm AF}$ is parallel to the $c$-axis, although the magnetization easy axis in the paramagnetic region is the $a$-axis.  
If the exchange interaction is isotropic, the ordered moment in the AFM state is expected to be along the $a$-axis. 
However, the experimental result is not the case.
CeAgSb$_2$ is a compound where the direction of the ordered magnetic moment is not the same as that in the paramagnetic region.\cite{CeAgSb2}
In order to clarify this unusual properties in CeAgSb$_2$, inelastic neutron scattering experiments were carried out and the existence of a very large anisotropic exchange interaction was revealed from the magnon dispersion. 
In fact, the unusual macroscopic properties in CeAgSb$_2$ were explained by considering the anisotropic exchange interaction. 
However, the origin of the anisotropic exchange interaction is not known. 
In the present compound, the anisotropic exchange interaction should be considered to explain the different direction of the ordered moment from the easy magnetization axis in the paramagnetic region. 
The inelastic neutron scattering experiments are necessary to know the magnon dispersion.
Here, we note the relation between the large spin gap observed in the inelastic neutron scattering experiments and the AFM order.
The $M$-$H$ curves for $H\parallel a$- and $c$-axes also suggest the small magnetic anisotropy in the AFM ordered phase from the nonexistence of the metamagnetic transition for $H\parallel a$- and $c$-axes. 
Thus, the magnon gap is expected to be small.  
On the other hand, experimental results (7), (8), and (9) strongly suggest the existence of a large spin gap.  
In a simple AFM compound with a large spin gap, a metamagnetic transition is expected to appear.  
However, it is not seen in the experiments.  
The relation between the small magnetic anisotropy in the AFM ordered phase and the existence of a large spin gap is not known at present. 
Clarifying the origin of a large spin gap is crucial for understanding the unusual AFM order in this compound.

Thus, the AFM order in this compound is not understood as a simple AFM one. 
At present, we do not know the answer. 
We could not succeed in obtaining information on the CEF ground state from the $M$-$H$ curve for $H\parallel c$-axis because $H_c^p$ is higher than 55 T. 
It is necessary to clarify the anisotropic magnetic phase diagram to understand the LRO in this compound. 
We have to measure the high-field magnetization in La-doped compounds where $H_c^p$ is expected to be small. 
If $H_c^p$ becomes smaller by La doping, it is possible to know how large is the anisotropy of the magnetic phase diagram and the magnitude of $M$ at $H_c^p$ and the slope of the $M$-$H$ curve above $H_c^p$ which induces important information on the CEF ground state.

Next, we mention the electronic properties of the ground state. 
The electrical resistivity $\rho$ does not show a metallic behavior but a Kondo semiconductor like behavior at high temperatures. 
$\rho$ exhibits a large increase just below $T_0$ and a metallic behavior with decreasing temperature below $\sim5$ K after showing a maximum  at $\sim23$ K.\cite{Nishi} 
The temperature dependence of $\rho$ at high temperatures suggests that the density of states at Fermi energy decreases with decreasing temperature. 
Namely, the pseudo gap is opened on the Fermi surface. 
Altough the temperature dependence of $\rho$ along all the crystal axes suggests a gap opening on the Fermi surface, we do not  know if a charge gap is opened on the Fermi surface or not at $T_0$ because the Hall resistivity does not necessarily shows a clear increase just below $T_0$ in magnetic fields along all the crystal axes.\cite{Tany2} 
The metallic behavior of $\rho$ below $\sim5$ K together with the $\gamma T$ term in the specific heat suggests the existence of the Fermi surface even below $T_0$. 
However, this is not direct evidence of the Fermi surface. 
The observation of the SdH oscillations in the present study first provides direct evidence of the existence of the Fermi surface. 
This result has significance for understanding the mechanism of the LRO. 

Finally, we note the following. 
In CeRu$_2$Al$_{10}$, the valence of the Ce ion is close to +3, and we discussed the present results from this stand point. 
However, by considering that a similar transition also exists in CeOs$_2$Al$_{10}$, which has a more valence fluctuating character than CeRu$_2$Al$_{10}$, the transition is considered to appear in a boundary region between the localized and itinerant characters, although the microscopic mechanism is not yet known.

In summary, we examined the magnetization and longitudinal magnetoresistance of CeRu$_2$Al$_{10}$ up to $\sim55$ T for $H\parallel$ $c$-axis. 
First, we pointed out the mystery of the AFM order in CeRu$_2$Al$_{10}$ that the direction of the $M_{\rm AF}\parallel c$-axis proposed by Khalyavin $et\ al$. is different from the $a$-axis expected from $\chi_a$$>$$\chi_c$$>$$\chi_b$ in the paramagnetic phase. 
$M$ shows $H$-linear increase up to $\sim55$ T, which indicates a larger critical field for the paramagnetic phase than $\sim50$ T for $H\parallel a$-axis. 
The $M$-$H$ curves for $H\parallel a$- and $c$-axes behave as if the magnetic anisotropy in the AFM phase is small, although the magnetic anisotropy in the paramagnetic phase is large. 
The Van Vleck term in the AFM phase is unusually large. 
These results indicate that the AFM order in CeRu$_2$Al$_{10}$ is not a  simple AFM one. 
In the magnetoresistance measurement, we observed the SdH oscillations below 4.2 K, from which the cross section of the Fermi surface normal to the $c$-axis is estimated to be $\sim1.0\times 10^{14}$ cm$^{-2}$.  
This is the first direct evidence of the existence of the Fermi surface in the ground state of the LRO.

\end{document}